\documentstyle[prl,aps,epsf]{revtex}
\newcommand{\newc}{\newcommand}
\newc{\beq}    {\begin{equation}}
\newc{\eeq}    {\end{equation}}
\newc{\beqa}    {\begin{eqnarray}}
\newc{\eeqa}    {\end{eqnarray}}

\def\PL{{\em Phys. Lett.}}
\def\PRL{{\em Phys. Rev. Lett. }}
\def\PRB{{\em Phys. Rev.} {\bf B} }

\def\PREP{{\em Phys. Rep.}  }

\begin{document}

\draft
\twocolumn[\hsize\textwidth\columnwidth\hsize\csname
@twocolumnfalse\endcsname

%\preprint{ cond-mat/000000}
\title{Calculation of Magnetic Penetration Depth Length
$\mbox{\boldmath $\lambda(T)$}$ in High Tc Superconductors}
\author{ Jae-Weon Lee$^1$, In-Ho Lee$^2$, and Sang Boo Nam$^1$\cite{sbnam}  \\}
\address{$^1$ Superconductivity Group, Korea Research Institute of Standards and Science,\\
   \it  Doryong-dong~ 1, Yusung-ku, Taejeon 305-600 Korea\\
 $^2$Korea  Institute for Advanced Study, Cheong Ryang Ri, DongdaeMoon Ku, Seoul 130-012, Korea.}
%\date{\today}
\maketitle
\begin{abstract}
The notion of  a finite pairing interaction energy range via Nam, results in the
 incomplete condensation in which not all states participate in pairings. The
states not participating in pairings are shown to yield the low
energy states responsible  for the  linear T dependence of
superelectron density at low T in a s-wave superconductor. We
present  extensive  quantitative calculations of $\lambda$(T) for
all T ranges, in good agreements with experiments. It is not
necessary to have nodes in the order parameter, to account for the
linear T dependence of $\lambda$(T) at low T in high Tc
superconductors.

\end{abstract}
\pacs{PACS: 74.72.-h, 74.25.Ha, 74.25.Nf}
%\hskip 3.8cm
%y\vskip2pc
]

%\section {Introduction}
One of crucial parameters in a superconductor is the magnetic
penetration depth length $\lambda(T) $ which reflects the
condensation carrier density, superelectron density $ \rho_s (T)$,
in the London model as \beq {\rho_s (T)}/{\rho_s(0)}= \left[
{\lambda(0)}/{\lambda(T)} \right ] ^2. \eeq

The $\rho_s(T)$ plays an important role for understanding the
nature of condensation. In the Gorter and Casimir two fluid model
(GC), $\rho_s(T)$ varies as $1-(T/T_c)^4$. But  the
BCS-$\rho_s(T)$ has an activation form at low $T$ via the order
parameter $ \Delta$ which indicates the excitation energy gap. The
measurements \cite{ybco,ybco2,bi,hg,la,sr} of $\lambda(T)$ at low
$T$ in high $T_c$ superconductors (HTS) are compatible with
neither the BCS result nor the GC picture. Data indicate the
linear $T$ dependence of $\rho_s(T)$ at low $T$. This linear $T$
dependence of $\lambda(T)$ in fact is taken as providing evidence
that the order parameter has nodes, suggesting the d-wave pairing
state\cite{scalapino}. On the other hand, one of us
\cite{nam1,nam2} has shown that the notion of a finite pairing
interaction energy range $T_d$ results in the incomplete
condensation and the low energy states responsible for the linear
$T$ dependence of $\rho_s(T)$ at low $T$ in a  s-wave
superconductor\cite{nam3}. Moreover, the incomplete condensation
yields the multi-connected superconductors(MS)\cite{ms} which can
account for the $\pi$-phase shift in Pb-YBCO SQUID\cite{pi} and
1/2 fluxoid quantum in the YBCO ring with  three grain boundary
junctions\cite{fluxoid}.

Recently, the oxygen isotope effect \cite{hoffer}, $T_c \propto M^{-\alpha}$ with
 $\alpha=0.4 \sim 0.49$
in LSCO single crystal, suggests the electron-phonon interaction
would play an important role for  understanding superconductivity
in cuprate materials. And the BSCCO bicrystal c-asxis twist
Josephson junction experiment\cite{Li} indicates the dominant
order parameter contains the s-wave and not d-wave component.
Moreover, no node in the order parameter is observed in the
angular dependence of the non-linear transverse magnetic moment of
YBCO in the Meissner state\cite{bhatt}. On the other hand, the
scanning tunneling microscope imaging the effects of individual
zinc impurity atoms on superconductivity in BSCCO \cite{pan} shows
the four fold symmetric quasiparticle cloud, indicating the d-wave
component. But no four fold is observed in the same system
\cite{4fold}. Perhaps, the observation of \cite{pan} may be a
reflection of the Fermi surface.

It is highly desirable to carry out
quantitative calculations of $\lambda(T)$ for all
$T$ ranges to see the accountability of finite $T_d$  picture for $\lambda(T)$
data of HTS.
In this letter, we present extensive quantitative calculations
of $\lambda(T)$ for all $T$ ranges in good agreements  with data of HTS.
For this, it is worthy to recapitulate  the pertinent results for the notion of
a finite $T_d$\cite{nam1}.

\begin{figure}[Fig0112]
\centering
% \hskip 1.5cm
%\leavevmode % unseen
\epsfysize=7cm \epsfbox{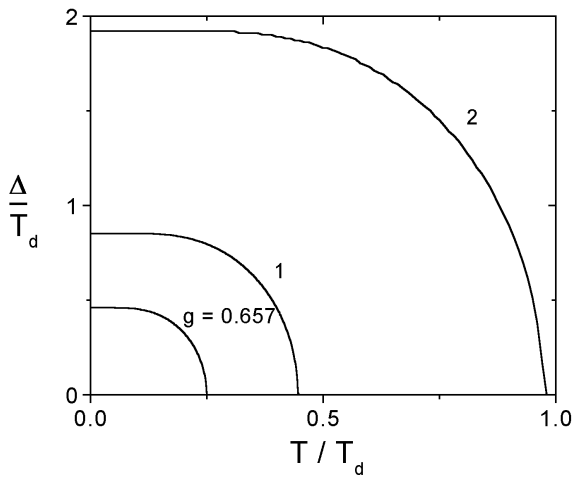}

\caption[Fig2]{\label{gap} The order parameter $\Delta$ as a function
of $g$ and $T$.}
\end{figure}

To see the phase transition, the transition temperature $T_c$
should be a finite value, that is, neither zero nor infinite. To
have a finite value of $T_c$, the pairing interaction energy range
$T_d$ should be finite, since $T_c$ is scaled with $T_d$ within
the pairing theory\cite{nam1}. In other words, the order parameter
$\Delta(k,\omega)$ may be written as\cite{nam1,nam2}

%\begin{eqnarray}
\beqa
\Delta(k, \omega)=\left\{
 \begin{array}{ll}
\Delta  &{\rm~~~for}~~ |\epsilon_k| <T_d \\ 0 &{\rm~~~for}~~  |\epsilon_k| > T_d
\end{array}
\right\}
\label{delta}
\eeqa
%\end{eqnarray}
for all frequencies $\omega$.
Here $\epsilon_k$ is the usual normal state excitation energy with the momentum $k$,
measured with respect to the Fermi level.
\begin{figure}[Fig0111]
% \hskip 1cm
\epsfysize=7cm \epsfbox{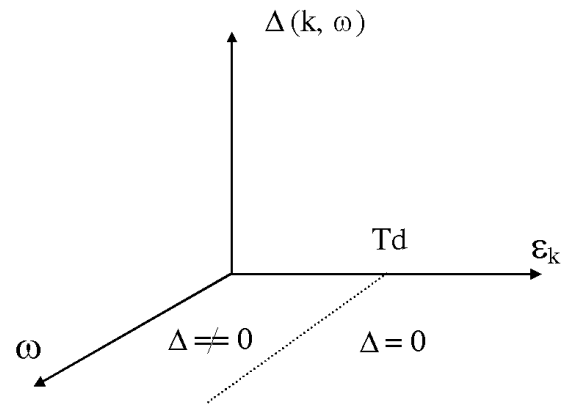}
%\
\caption[Fig1]{\label{phase}Schematic diagram showing
 the order parameter $\Delta$ in the finite pairing interaction energy ranges $T_d$ view.
}
\end{figure}
Here the natural units of $\hbar = c = k_B = 1$ are used. Later we
use $\Delta(T)$ with $T$ for $\Delta$ as well. Note that $w$ has
no constraint and that  in the case of pairings of carriers via
the electron-phonon interaction, the $T_d$ corresponds to the
Debye temperature. However, the nature of $T_d$ in HTS is still
unknown. Our results are not depending on the nature of $T_d$. The
order parameter $\Delta$  is the solution of the BCS like equation
\cite{nam1}  \beq \frac{1}{g}=\int^{T_d}_0 \frac{d \epsilon}{E} ~
tanh \frac{\beta}{2} E, \label{bcs} \eeq where $E=(\epsilon^2
+\Delta^2)^{1/2}, \beta=1/T$, and $g$ corresponds to the BCS
coupling parameter $N(0) V_{BCS}$. The solution of $\Delta$  for
$g$ are shown in the unit of $T_d$ in Fig.~\ref{gap}.

\begin{figure}[Fig0113]
% \hskip 1.5cm
\epsfysize=7cm \epsfbox{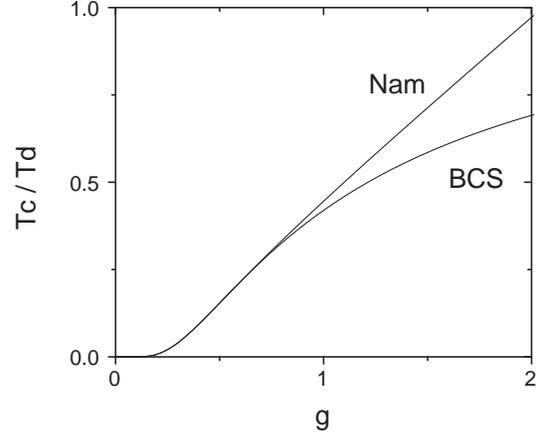}
\
\caption[Fig3]{\label{bcsfig} $T_c$ versus $g$ for BCS and Nam's model\cite{nam1}, respectively.
}
\end{figure}

The equation for $T_c$ from Eq.~(\ref{bcs})  via $\Delta(T_c)=0$
may be written as\cite{nam1,nam2} \beq 1/g=(2/\pi)\sum_j (2/j)
tan^{-1} (y/j), \label{g2} \eeq where $y=T_d/\pi T_c$, and sum is
over  the positive odd integers $j$. The factor of arctangent
function makes the sum converge. For large $y$, Eq.~(\ref{bcs})
yields the BCS result $T_c$(BCS). The quantitative calculations of
$T_c$ are given in Fig.~\ref{bcsfig}, together with the BCS
$T_c(BCS)$. Unlike the BCS result of $T_c(BCS)$, the $T_c$ from
Eq.(\ref{g2}) does not have any upper limit. The fact is that for
large $g > 2.32$, $T_c$ increases with increasing g as $T_c =
gT_d/2$. One interesting value of g = 0.657 yields $T_c$ = 100 K
with $T_d$ = 400 K which is of the order of the Debye temperature
in HTS. This value of g may be realized in YBCO by considering the
electron-phonon interaction of the order of $\lambda_p = 1.3 \sim
2.3$ \cite{weber}.

The BCS parameter $\Delta(0)/T_c$ can  be easily calculated from
Eq.~(\ref{bcs})  as \cite{nam1}

\beq \Delta(0)/T_c=T_d/[T_c ~ sinh(1/g)]. \eeq

\begin{figure}[Fig0113]
% \hskip 1.5cm
\epsfysize=7cm \epsfbox{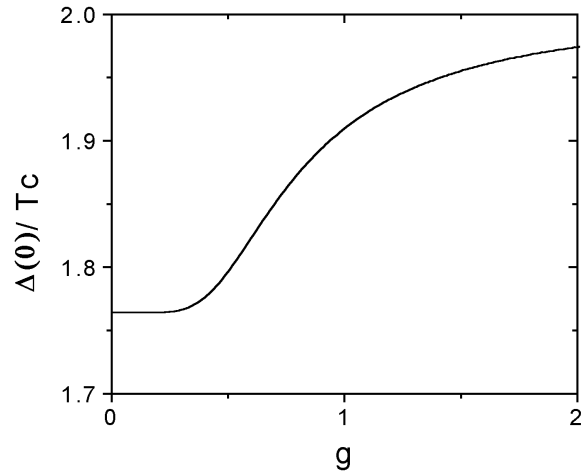}
\
\caption[Fig3]{\label{delta0} The BCS parameter $\Delta(0)/T_c$ versus $g$.
}
\end{figure}

As is shown in Fig.~\ref{delta0}, $\Delta(0)/T_c$ is  a  function
of $g$ or $T_c/T_d$, and  increases with increasing $g$ or
$T_c/T_d$. In the range of $g <0.2$ or $T_c/T_d < 0.0076$, it has
a constant BCS value. In fact, this range corresponds to the case
of  low $T_c$ superconductors(LTS). In the range of $0.5 < g< 1.5$
it increases almost in a linear of $g$, and has a  saturated value
of $2$ for large $g$.

In the sprit of Bardeen\cite{bardeen}, the normal fluid density
$\rho_n(T)=\rho-\rho_s(T)$, within the pairing theory, may be
written as \cite{scalapino} \beq \rho_n(T)/\rho=2\int^{\infty}_0
d(\omega/T) n(\omega) f(\omega/T) \left [ 1- f(\omega/T) \right ],
\label{rho} \eeq where $f(x)$  is the usual Fermi function
$1/[1+exp(x)]$ and the density of states
$n(\omega)=N(\omega)/N(0)$ is given by \cite{nam1}
\begin{eqnarray}
n(\omega) &=& q(\omega/T_d)+n_{\rm BCS}(\omega)r(\omega/T_d),\\
\nonumber \\ \label{q} q(\omega/T_d) &=&
(2/\pi)\tan^{-1}(\omega/T_d),\\ \nonumber\\ r(\omega/T_d) &=&
(2/\pi)\tan^{-1}[n_{\rm BCS}(\omega)T_d /\omega],\\ \nonumber\\
n_{\rm BCS}(\omega) &=& Re \{\omega/(\omega^2-\Delta^2)^{1/2}\}.
\end{eqnarray}
Physically, $\rho_n(T)$ would be resulted from the single particle
excitation not
 pairs.
Thus, the factor $f(x)$ in Eq.~(\ref{rho}) is the occupation
probability of the state $|k\uparrow>$ and the factor $[1- f(x)]$
is the unoccupation probability of the partner state, say,
$|-k\downarrow>$ , and vice versa, respectively. The factor $2$
comes from the spin sum. The states  of Eq.~(\ref{q})  are
reflections of states being not participated in pairings. A word
of caution is in order. The $\omega$ is the dynamical energy which
has the kinetic as well as interaction parts. Physically, the sum
of spectral weights outside $T_d < |\epsilon_k|$, result in the
states of Eq.~(\ref{q}). Thus, the low energy states are realized.
In other words, carriers which do not participate in pairings
yield the linear T dependence of $\rho_n(T)$ at low $T$. In fact,
these states results in the linear $T$ dependence of $\lambda(T)$
at low T. To see this, by inserting Eq.~(\ref{q}) into
Eq.~(\ref{rho}), one can get the variation of $\lambda(T)$ at low
$T$ as \cite{nam2,nam3}
\begin{eqnarray}
\Delta\lambda/\lambda(0) = \frac{1}{2} \rho_n(T)/\rho =(T/T_c) (T_c/T_d)(2/\pi)\ln 2,
\label{slope}
\end{eqnarray}
similar to the result by d-wave picture\cite{scalapino},
\begin{eqnarray}
[\Delta \lambda/\lambda(0)]_d = (T/T_c)(T_c/\Delta_0) \ln 2
\end{eqnarray}
via $n_d(\omega) = \omega/\Delta_0$, where $\Delta_0$ is the maximum value (anti-node) of the order parameter.

For the quantitative calculations of $\lambda(T)$,
we have determined $T_c/T_d$ or $g$~[Eq.~(\ref{g2})]
via Eq.~(\ref{slope}), by taking the slope of $[\lambda(0)/\lambda(T)]^2$ near
zero  temperature. Once $g$  or $T_c/T_d$ is
set, no adjustable parameter is used in our calculations of Eq.~(\ref{rho}).

\begin{figure}[Fig0115]
% \hskip 1.5cm
\epsfysize=7cm \epsfbox{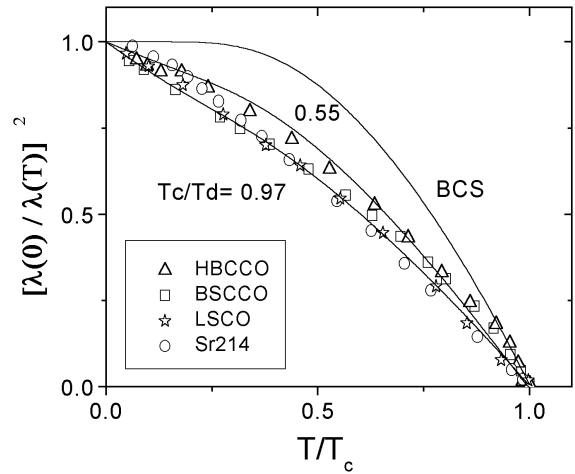}
\
\caption[Fig1]{\label{lambda1} The  temperature
dependence of $[\lambda(0)/\lambda(T)]^2$ (solid lines) compared with
the experimental data for HBCCO\cite{hg}, BSCCO\cite{bi}, LSCO \cite{la} and Sr214\cite{sr}.}
\end{figure}

As is shown in Fig.~\ref{lambda1}, we  have obtained good agreements between calculations and data of
 BSCCO by Lee
et al\cite{bi}, HBCCO by Panagopoulos et al\cite{hg}
 and LSCO by Panagopoulos et al\cite{la}, and Sr214 by Bonalde et al\cite{sr}, respectively.
 We picked up not all of data points in the papers for clarity.
 Bonalde et al \cite{sr} reported that their data at low $T$ vary as
 $T^2$ which are resulted from scatterings by impurities or
 defects. In a finite $T_d$ picture, the impurity scatterings
 make some states at the Fermi level not participate in pairings,
 and result in the $T^2$ term in $\lambda(T)$ at low T
 \cite{nam4}.

\begin{figure}[Fig0116]
% \hskip 1.5cm
\epsfysize=7cm \epsfbox{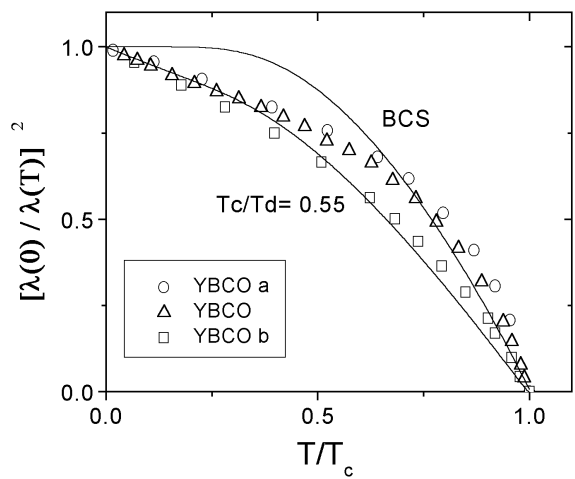}
\
\caption[Fig1]{\label{lambda2} $[\lambda(0)/\lambda(T)]^2$ (solid line)
compared with data for bulk YBCO \cite{ybco}, and
 a- and  b-axes, respectively \cite{ybco2}.
}
\end{figure}

However, as  shown in Fig.~\ref{lambda2}, we have obtained poor
agreement near $T_c$ between calculation and data of YBCO by Hardy
et al \cite{ybco} and anisotropic data of YBCO by Kamal et
al\cite{ybco2}. The YBCO b case is good.

The notion of a finite $T_d$ results in the incomplete condensation
at zero temperature.
By considering the sum rule, we can calculate the fraction of states, $R$,
being  not participated in pairings as \cite{nam1}
\beqa
R(z)&=&\int^{\infty}_\Delta \left [ n_{BCS} (\omega) - n(w) \right ] d\omega /\Delta \\
     &=& \int^{\Delta}_{0} n(\omega) d \omega /\Delta \nonumber \\
&=&(2/\pi)tan^{-1}(z) -(1/z\pi) ln(1+z^2) \nonumber,
\eeqa
where $ z=\Delta/T_d=[\Delta(0)/T_c](T_c/T_d)$ which is  a
function of $g$ or $T_c/ T_d$.

\begin{figure}[Fig0111]
% \hskip 1.5cm
\epsfysize=7cm \epsfbox{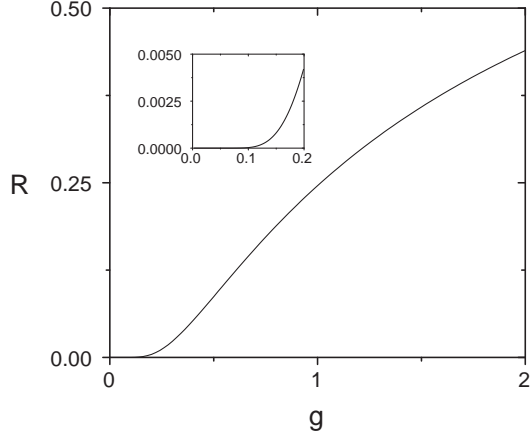}
\
\caption[Fig5]{\label{R}
 The fraction of states, $R$,
being  not participated in pairings versus $g$.}
\end{figure}

In Fig.~\ref{R} is shown $R(z)$ as a function of $g$.
In the range of $g < 0.2$ or $T_c/T_d <0.0076$, $R(z)$ is negligible.
As stated before,
this range corresponds
to the case of LTS. Thus, the linear $T$ dependence
of $\lambda(T)$ at low $T$ is hardly observed in LTS.

In summary,  even though the model of Eq.~(\ref{delta})
is  ideal, the quantitative calculations account  very well for data for
  all $T$ ranges without any adjustable parameter, except for YBCO data near
$T_c$. Perhaps, the Fermi surface effect would play an important
role for $\lambda(T)$ in the case of YBCO. Of course, the
retardation and non-local effects  should be taken into  account
as well for improvement. In all, the calculations are quite
satisfactory and theoretically sound. We suggest the pairing
interaction energy range $T_d$ in HTS may be of the order of $1
\sim 2 $ times $T_c$. The linear $T$ dependence  of $\lambda(T)$
at low $T$ does not imply nodes in the order  parameter, contrary
to general belief.

In the spirit of  a finite $T_d$, it is recently shown\cite{nam4}
that  the spinless impurity scatterings suppress $T_c$ and destroy
superconductivity. Some states at the Fermi level are shown not to
participate in pairings when there are scattering centers such as
impurities, and result in the linear $T$ term in the specific heat
at low $T$. The quantitative calculations\cite{ilee}
 account well for the reduction of Tc \cite{xiao} and the specific heat data \cite{moler,sisson} in the Zn-doped YBCO, respectively.

%\subsection*{}
%\vskip{1cm}
We thank KRISS members for their warm hospitalities at KRISS.
Specially JWL thanks Dr. Y. H. Lee for his kindness and SBN thanks
Drs J. C. Park and Y. K. Park for various discussions. This work
is supported in part by KOFST.

\end{document}